\documentclass[dvips,draft,nolinenumbers]{ectaart}

\usepackage[utf8]{inputenc}
\usepackage{amsmath}
\usepackage{amsfonts}
\usepackage{multirow}
\usepackage{enumerate}
\usepackage{natbib}
\usepackage[colorlinks=true,linkcolor=black,citecolor=black,urlcolor=black]{hyperref}
\usepackage{color}

\startlocaldefs
\newtheorem{example}{Example}
\newtheorem{theorem}{Theorem}
\endlocaldefs

\begin{document}

\begin{frontmatter}

\title{Overcoming the inconsistences of the variance inflation factor: a redefined VIF and a test to detect statistical troubling multicollinearity. }
\runtitle{A redefined VIF and a test to detect statistical troubling multicollinearity.}


\begin{aug}
\author{\fnms{Rom\'an} \snm{Salmer\'on-G\'omez}\ead[label=e1]{Dept. of Quantitative Methods for Economics and Business. University of Granada.}}
\address{\printead{e1}}
\author{\fnms{Catalina B.} \snm{Garc\'ia-Garc\'ia}\ead[label=e1]{Dept. of Quantitative Methods for Economics and Business. University of Granada.}}
\address{\printead{e1}}
\author{\fnms{Jos\'e} \snm{Garc\'ia-P\'erez}\ead[label=e2]{Dept. of Economics and Business. University of AlmerÃ­a.}}
\address{\printead{e2}}
\end{aug}

\runauthor{Salmer\'on et al.}

\begin{abstract}
Multicollinearity is relevant to many different fields where linear regression models are applied, and its existence may affect the analysis of ordinary least squares (OLS) estimators from both the numerical and statistical points of views. Thus, multicollinearity can lead to incoherence in the statistical significance of the independent variables and the global significance of the model. The variance inflation factor (VIF) is traditionally applied to diagnose the possible existence of multicollinearity, but it is not always the case that detection by VIF of a troubling degree of multicollinearity corresponds to negative effects on the statistical analysis. The reason for the lack of specificity of VIF is that there are other factors, such as the size of the sample and the variance of the random disturbance, that can lead to high values of the VIF but not to problematic variance in the OLS estimators (see \cite{OBrien}). This paper presents a new variance inflation factor (TVIF) that consider all these additional factors. Thresholds for this new measure and from the index provided by Stewart \cite{Stewart:1987} are also provided. These thresholds are reinterpreted and presented as a new statistical test to diagnose the existence of statistical troubling multicollinearity. The contributions of this paper are illustrated with two real data examples previously applied in the scientific literature.
\end{abstract}


\end{frontmatter}

\section{Introduction}

The existence of near-troubling multicollinearity in a linear regression model is caused by a strong linear relationship between at least two independent variables and may affect the analysis of the ordinary least squares (OLS) estimators from both the numerical and statistical point of views. Focusing on the second case, obtaining an inflated estimated variance leads to the tendency not to reject the null hypothesis in the individual significance test while, at the same, rejecting the null hypothesis of the global significance test. This is an incoherence that is commonly identified as a symptom of near-multicollinearity. However, the measures traditionally applied to detect multicollinearity may conclude that multicollinearity exists even if it does not lead to the negative effects mentioned above, thereby suggesting that best solution  is not to treat the multicollinearity (see \cite{OBrien}).

This paper proposes an alternative procedure focusing on checking if the detected multicollinearity affects the statistical analysis of the model. For this focus on disruption, it measures are needed to indicate if the statistical analysis of the model is affected by the existing near-multicollinearity; introducing such a measure is the main goal of this paper.

Thus, since the near-multicollinearity is caused by a linear relationship between the independent variables in the model, a natural way to measure this relation is from the coefficient of determination of the auxiliary regression of each independent variable as a function of the rest of the independent variables of the model. It is commonly accepted practice to conclude that multicollinearity exists if this coefficient of determination is high, with 0.9 being the commonly accepted threshold.

The variance inflation factor (VIF) is obtained from this coefficient of determination; consequently, the VIF is able to diagnose the degree of near-multicollinearity existing in the model. Thus, there is a VIF for each independent variable except for the intercept, for which it is not possible to calculate a coefficient of determination for the corresponding auxiliary regression.

If the VIF is high (with 10 being the traditionally accepted threshold), it is said that the multicollinearity is troubling. \cite{Salmeron2018} and \cite{Salmeron2019} showed that this measure is only able to diagnose the essential near-multicollinearity (i.e., the relationships between the independent variables of the model except for the intercept, see \cite{MarquardtSnee}).

The VIF is also defined (see, for example, \cite{Johnston1984}) as the ratio between the variance of the OLS estimator of the original model and the variance of the model in which the variables are orthogonal. Indeed, the name of this measure comes from this idea due to how it quantifies how the variance of the OLS estimators have increased on account of the near-multicollinearity existing in the model. This inflation also implies a diminishing of the experimental value of the individual significance test, leading to a tendency not to reject the null hypothesis of this type of test.

However, as previously noted, not every instance in which VIF detects a troubling degree of multicollinearity corresponds to negative effects on the statistical analysis. Such a lack of specificity results from the fact that other factors, such as the size of the sample and the variance of the random disturbance, can lead to high values of the VIF but not increase the variance of the OLS estimators (see \cite{OBrien}). The explanation for this phenomenon hinges on the fact that, in the model with orthogonal variable that is traditionally considered as the reference, the linear relations are assumed to be eliminated whereas other factors, such as the variance of the random disturbance, maintain the same values. Although \cite{Johnston1984} stated that ``the orthogonal case does not mean that it is a realizable goal but it is used as a reference point to measure the relative increasing of the sample variance of the estimators'', we consider that it should be a plausible goal. For this, the first step is to establish a reference orthogonal model that permits an analysis not only of how much the VIF is altered but also of how much the other relevant factors in the model are altered.

Then, this paper proposes a QR decomposition in the matrix of independent variables of the model in order to obtain a orthonormal matrix. By redefining the reference point, the variance inflation factor is also redefined, yielding a new measure of detection that analyzes the alteration of the VIF and the rest of relevant factors of the model, thereby overcoming the problems associated with the traditional VIF, as described by \cite{OBrien} among others. Thus, the intercept is included in the detection (in contrast to what occurs with the traditional VIF); Of even greater relevance, this new measure is defined as a statistical test for detecting the troubling near-multicollinearity. Note that most measures used to diagnose multicollinearity are merely indicators with rules of thumb rather than statistical tests per se. To the best of our knowledge, the only extant statistical test for the diagnosis of multicollinearity was presented by \cite{FarrarGlauber} and has received strong criticism (see, for example, \cite{CriticaFarrar1}, \cite{CriticaFarrar2}, \cite{CriticaFarrar3} and  \cite{CriticaFarrar4}). Thus, the statistical test presented in this paper should be a relevant contribution in the field of econometrics insofar as it would fill an existing gap in the scientific literature.

The paper is structured as follows: Section \ref{preliminares} presents some preliminaries to introduce the methodology applied to establish the mentioned test. Section \ref{modelo.orto} proposes considering as a reference an alternative orthogonal model from which obtain lower bounds for the VIF to determine if the degree of near-multicollinearity existing in the model is affecting its statistical analysis. These bounds are reinterpreted and presented as a statistical test. The alternative orthogonal model leads to a new definition of the VIF and its thresholds, as presented in Section \ref{new.VIF.orto}, and new thresholds for the index introduced by Stewart \cite{Stewart:1987}, as presented in Section \ref{cotas.ki.apartiTVIFR}. Finally, Section \ref{conclusiones.VIF} summarizes the main contributions of this paper.

\section{Preliminaries}
    \label{preliminares}

By considering the following multiple linear regression model with $n$ observations and $k$ independent variables
\begin{equation}
    \mathbf{y}_{n \times 1} = \mathbf{X}_{n \times k} \cdot \boldsymbol{\beta}_{k \times 1} + \mathbf{u}_{n \times 1},
    \label{model0}
\end{equation}
the VIF is one of the most commonly applied measures for diagnosing troubling near-multicollinearity. This measure is traditionally defined as the ratio between the variance of the estimator in the original model, $var \left( \widehat{\beta}_{i} \right)$, with the variance of the estimator of a model in which is orthogonality is presupposed among the independent variables, $var \left( \widehat{\beta}_{i,o} \right)$. This is to say:
\begin{equation}
    \label{vari.VIF}
    var \left( \widehat{\beta}_{i} \right) = \frac{\sigma^{2}}{n \cdot var(\mathbf{X}_{i})} \cdot \frac{1}{1 - R_{i}^{2}} =  var \left( \widehat{\beta}_{i,o} \right) \cdot VIF(i), \quad i=2,\dots,k,
\end{equation}
where $\mathbf{X}_{i}$ is the  independent variable $i$ of the model  (\ref{model0}), and $R^{2}_{i}$ the coefficient of determination of the following auxiliary regression:
\begin{equation}
    \mathbf{X}_{i} = \mathbf{X}_{-i} \cdot \boldsymbol{\alpha} + \mathbf{v},
    \label{model_aux}
\end{equation}
where $\mathbf{X}_{-i}$ is the result obtained when  $\mathbf{X}_{i}$ is eliminated from matrix $\mathbf{X}$.

As observed in expression (\ref{vari.VIF}), a high VIF leads to a high estimated variance; then, the experimental value for the individual significance test given by
\begin{equation}
    \label{texp.orig}
    t_{exp}(\beta_{i}) = \left| \frac{\widehat{\beta}_{i}}{\sqrt{\frac{\widehat{\sigma}^{2}}{n \cdot var(\mathbf{X}_{i})} \cdot VIF(i)}} \right|, \quad i=2,\dots,k,
\end{equation}
will be low, provoking the tendency not to reject the null hypothesis because the criterion to reject the null hypothesis is not met, i.e., the experimental statistic will be greater than the theoretical statistic (given by $t_{n-k}(1-\alpha/2)$, where $\alpha$ represents the significance level).

However, this statement is full of simplifications. By following  \cite{OBrien}, and as is easily observed in expression (\ref{texp.orig}), other factors, such as the estimation of the random disturbance and the size of the sample, can counterbalance the high value of the VIF to yield a low value for the experimental statistic.  That is to say, it is possible to obtain VIF values greater than 10 (the threshold traditionally established to determine troublesomeness) that do not necessarily imply high estimated variance on account of a large sample size or a low value for the estimated variance of the random disturbance. This explains why not all occasions with a high value for the VIF see effects on the statistical analysis of the overall model.

By considering the initial orthogonal model, the value of the experimental statistic of the individual significance test whose null hypothesis is $\beta_{i} = 0$ in the face of the alternative hypothesis $\beta_{i} \not= 0$, with $i=2,\dots,k$ will be given by
\begin{equation}
    \label{texp.orto.1}
    t_{exp,O}(\beta_{i}) = \left| \frac{\widehat{\beta}_{i}}{\sqrt{\frac{\widehat{\sigma}^{2}}{n \cdot var(\mathbf{X}_{i})}}} \right|, \quad i=2,\dots,k,
\end{equation}
where the estimated variance has been diminished due to the VIF always being greater than or equal to 1, and consequently,  $t_{exp,O}(\beta_{i}) > t_{exp}(\beta_{i})$. However, it has been supposed that the same estimations are obtained in the orthogonal and the original models, which does not seem a plausible supposition.

On the other hand, to determine if the tendency not to reject the null hypothesis in the individual significance test is provoked by a troubling near-multicollinearity that is inflating the variance of the estimator versus being due to both variables’ not being significatively related from an statistical point of view, the following situations are distinguished:
\begin{enumerate}[a)]
    \item If the null hypothesis is initially rejected, $t_{exp}(\beta_{i}) > t_{n-k}(1-\alpha/2)$, it is expected that it will be also rejected in the orthogonal case, although this is not assured. Proof of this inconsistency will be given in subsequent sections.
    \item If the null hypothesis is not initially rejected, $t_{exp}(\beta_{i}) < t_{n-k}(1-\alpha/2)$, the following findings for the orthogonal model can occur:
        \begin{enumerate}[b.1)]
            \item the null hypothesis is not rejected either, $t_{exp,O}(\beta_{i}) < t_{n-k}(1-\alpha/2)$; then, it will be possible to conclude that the degree of multicollinearity does not justify not rejecting the null hypothesis in the initial model
            \item the null hypothesis is rejected, $t_{exp,O}(\beta_{i}) > t_{n-k}(1-\alpha/2)$; then, it becomes possible to conclude that the degree of multicollinearity justifies not rejecting the null hypothesis in the initial model with regard to its affecting the statistical analysis of the model.
        \end{enumerate}
\end{enumerate}

Thus, by taking into account the expressions (\ref{texp.orig}) and (\ref{texp.orto.1}), it is verified that $t_{exp,O}(\beta_{i}) = t_{exp}(\beta_{i}) \cdot \sqrt{VIF(i)}$. Consequently, in the orthogonal case, the null hypothesis is rejected if
\begin{equation}
    VIF(i) > \left( \frac{t_{n-k}(1-\alpha/2)}{t_{exp}(\beta_{i})} \right)^{2} = c_{1}(i), \quad i=2,\dots,k.
\end{equation}
That is to say, if the VIF associated with the variable $i$ is greater than the upper bound $c_{1}(i)$ it can be concluded that the estimator of that variable is significatively different than zero in the hypothetical case in which the variables are orthogonal. If the null hypothesis is not rejected in the initial model, the reason for the failure to reject will be attributable to the degree of multicollinearity that is affecting the statistical analysis of the model.

Finally, note that since the interesting cases are those in which the null hypothesis is initially not rejected,  $t_{exp}(\beta_{i}) < t_{n-k}(1-\alpha/2)$, the upper bound $c_{1}(i)$  will always be greater than one.
    \begin{table}
        \begin{center}
        \begin{tabular}{ccccc}
            \hline
            Year & $\mathbf{D}$ & $\mathbf{C}$ & $\mathbf{I}$ & $\mathbf{CP}$  \\
            \hline
            1996 & 3.80510 & 4.7703 & 4.8786 & 808.23  \\
            1997 & 3.94580 & 4.7784 & 5.0510 & 798.03 \\
            1998 & 4.05790 & 4.9348 & 5.3620 & 806.12 \\
            1999 & 4.19130 & 5.0998 & 5.5585 & 865.65 \\
            2000 & 4.35850 & 5.2907 & 5.8425 & 997.30 \\
            2001 & 4.54530 & 5.4335 & 6.1523 & 1140.70\\
            2002 & 4.81490 & 5.6194 & 6.5206 & 1253.40\\
            2003 & 5.12860 & 5.8318 & 6.9151 & 1324.80 \\
            2004 & 5.61510 & 6.1258 & 7.4230 & 1420.50\\
            2005 & 6.22490 & 6.4386 & 7.8024 & 1532.10 \\
            2006 & 6.78640 & 6.7394 & 8.4297 & 1717.50\\
            2007 & 7.49440 & 6.9104 & 8.7241 & 1867.20 \\
            2008 & 8.39930 & 7.0993 & 8.8819 & 1974.10 \\
            2009 & 9.39510 & 7.2953 & 9.1636 & 2078.00 \\
            2010 & 10.68000 & 7.5614 & 9.7272 & 2191.30  \\
            2011 & 12.07100 & 7.8036 & 10.3010 & 2284.90\\
            2012 & 13.44821 & 8.0441 & 10.9830 & 2387.50 \\
            \hline
        \end{tabular}
        \caption{Data set presented previously by \cite{Wissell}} \label{datos.Wissel}
        \end{center}
    \end{table}

\begin{example}\nonumber
    \label{ejemplo1}

    Table \ref{datos.Wissel} shows a data set (previously presented by \cite{Wissell}) presenting outstanding mortgage debt ($\mathbf{D}$, trillions of dollars), personal consumption ($\mathbf{C}$, trillions of dollars), personal income ($\mathbf{I}$, trillions of dollars) and outstanding consumer credit ($\mathbf{CP}$, trillions of dollars) for the years 1996–2012. Table \ref{tabla.W.orto.1} shows the OLS estimation of the model explaining the outstanding mortgage debt as a function of the rest of the variables. Note that the estimations for the coefficients of personal consume, personal income and outstanding consumer credit are not significatively different from zero\footnote{A level of significance equal to 5\% is considered through the paper.}, while the model is considered to be globally valid. This is traditionally understood as an unequivocal symptom of statistical troubling multicollinearity.

    \begin{table}
        \begin{tabular}{cccc}
            \hline
            Variable & Estimator & Standard deviation & Experimental $t$ \\
            \hline
            Constant & 5.469 & 13.016 & 0.420 \\
            Personal consumption & -4.252 & 5.135 & 0.828 \\
            Personal income & 3.1203 & 2.035 & 1.533 \\
            Outstanding consumer credit & 0.0028 & 0.0057 & 0.499 \\
            \hline
            $\widehat{\sigma}^{2}$ & \multicolumn{3}{c}{$0.9325^{2}$} \\
            $R^{2}$ & \multicolumn{3}{c}{0.9235} \\
            $F_{3,13}$ & \multicolumn{3}{c}{52.3} \\
            \hline
        \end{tabular}
        \caption{OLS estimation for the data set presented in Table  \ref{datos.Wissel}} \label{tabla.W.orto.1}
    \end{table}

    Taking into account that $t_{13}(0.975) = 2.16037$, it is verified that $c_{1}(2) = 6.807627$, $c_{1}(3) = 1.985966$ and $c_{1}(4) = 18.7437$. Since the VIFs are equal to 589.754, 281.8862 and 189.4874, respectively, it is concluded that the individual significance values for the three cases are affected by the degree of multicollinearity existing in the model.\hfill $\Box$
\end{example}

However, in the orthogonal case, it will be verified that $\mathbf{X}^{t} \mathbf{X} = \mathbf{D}$, where $\mathbf{D} = diag(d_{1},\dots,d_{k})$ a diagonal matrix, and it has to be verified that $var \left( \widehat{\beta}_{i,o} \right) = \sigma^{2} \cdot d_{i}^{-1}$, with $d_{i} = \mathbf{X}_{i}^{t} \mathbf{X}_{i}$ for $i=1,\dots,k$. This expression differs from the one given in (\ref{vari.VIF}), due to
$$\frac{\sigma^{2}}{n \cdot var(\mathbf{X}_{i})} \not= \frac{\sigma^{2}}{\mathbf{X}_{i}^{t} \mathbf{X}_{i}} = \frac{\sigma^{2}}{n \cdot ( var(\mathbf{X}_{i}) + \overline{\mathbf{X}}_{i}^{2})},$$
which even assumes that the estimation of $\sigma^{2}$ overlaps between the original and orthogonal model (a fact that should be checked).
In conclusion, it seems that this orthogonal model is not the most adequate point of reference.

\section{A new test from an alternative orthogonal model} \label{modelo.orto}

This section introduces a statistical test departing from the rule presented in the previous section but proposing an alternative orthogonal model. First, this alternative orthogonal model is presented, and its utility as an adequate point of reference is demonstrated. Then, the statistical test is presented and illustrated with the same example  as used previously to show the relevant differences.

\subsection{An alternative orthogonal model}

From a QR decomposition of matrix $\mathbf{A}_{n \times k}$, a square orthonormal\footnote{An orthonormal matrix is a orthogonal matrix with unit length columns.} matrix $\mathbf{A}_{o}$ is obtained having the same dimensions as $\mathbf{A}$, as well as another superior triangular matrix $\mathbf{P}_{k \times k}$ such that $\mathbf{A} = \mathbf{A}_{o} \cdot \mathbf{P}$.
By applying this calculation to matrix $\mathbf{X}$ of model (\ref{model0}), matrices $\mathbf{X}_{o}$ and $\mathbf{P}$ are obtained that verify the previous conditions. In this case, a hypothetical orthonormal model $\mathbf{y} = \mathbf{X}_{o} \cdot \boldsymbol{\beta}_{o} + \mathbf{w}$ is obtained such that
\begin{eqnarray}
    \widehat{\boldsymbol{\beta}} &=& \left( \mathbf{X}^{t} \mathbf{X} \right)^{-1} \mathbf{X}^{t} \mathbf{y} = \left( \mathbf{P}^{t} \mathbf{X}_{o}^{t} \mathbf{X}_{o} \mathbf{P} \right)^{-1} \mathbf{P}^{t} \mathbf{X}_{o}^{t} \mathbf{y} \nonumber \\
        &=& \mathbf{P}^{-1} \cdot \left( \mathbf{X}_{o}^{t} \mathbf{X}_{o} \right)^{-1} \mathbf{X}^{t}_{o} \mathbf{y} = \mathbf{P}^{-1} \cdot \widehat{\boldsymbol{\beta}}_{o}, \nonumber \\
    \mathbf{e} &=& \mathbf{y} - \mathbf{X} \cdot \widehat{\boldsymbol{\beta}} = \mathbf{y} - \mathbf{X}_{o} \mathbf{P} \mathbf{P}^{-1} \cdot \widehat{\boldsymbol{\beta}}_{o} = \mathbf{y} - \mathbf{X}_{o} \cdot \widehat{\boldsymbol{\beta}}_{o} = \mathbf{e}_{o}, \nonumber \\
    var \left( \widehat{\boldsymbol{\beta}}_{o} \right) &=&  var \left( \mathbf{P} \cdot \widehat{\boldsymbol{\beta}} \right) = \mathbf{P} \cdot var \left( \widehat{\boldsymbol{\beta}} \right) \cdot \mathbf{P}^{t} = \sigma^{2} \cdot \mathbf{P} \left( \mathbf{X}^{t} \mathbf{X} \right)^{-1} \mathbf{P}^{t} \nonumber \\
        &=& \sigma^{2} \cdot \mathbf{P} \left( \mathbf{P}^{t} \mathbf{X}_{o}^{t} \mathbf{X}_{o} \mathbf{P} \right)^{-1} \mathbf{P}^{t} = \sigma^{2} \cdot \left( \mathbf{X}^{t}_{o} \mathbf{X}_{o} \right)^{-1} = \sigma^{2} \cdot \mathbf{I}. \nonumber
\end{eqnarray}

Due to $\mathbf{X}_{o}$ being an orthonormal matrix, it is verified that $\mathbf{X}^{t}_{o} \mathbf{X}_{o} = \mathbf{I}$, and in that case,  $var ( \widehat{\beta}_{i,o} ) = \sigma^{2}$. This is to say, $d_{i} = 1$ for all $i$. On the other hand, since the errors are the same in the original and orthonormal model (which is the same as saying that both models provide the same estimates for the dependent variable), the estimation of $\sigma^{2}$ is the same in both models ($n$ and $k$ are not altered\footnote{Indeed, as the dependent variable is the same, the total sum of squares (TSS) is also the same, as are, consequently, the coefficient of determination and the statistic of the globally significance test.}). Thus, taking into account expression \ref{vari.VIF} it will be possible to conclude that
\begin{equation}
    \label{redef.VIF}
    \frac{var \left( \widehat{\beta}_{i} \right)}{var \left( \widehat{\beta}_{i,o} \right)} = \frac{VIF(i)}{n \cdot var(\mathbf{X}_{i})}, \quad i=2,\dots,k,
\end{equation}
which differs from the one deduced from (\ref{vari.VIF}). Note that for  $var \left( \widehat{\beta}_{i,o} \right) < var \left( \widehat{\beta}_{i} \right)$ to be verified, it is necessary that $VIF(i) > n \cdot var(\mathbf{X}_{i})$ for $i=2,\dots,k$. Thus, it is not assured that the estimated variance of the estimators for the orthogonal model will be lower that the variances of the estimators for the initial model, due to it also depending on the size of the sample and the variances of each independent variable.

Finally, given the orthonormal model $\mathbf{y} = \mathbf{X}_{o} \cdot \boldsymbol{\beta}_{o} + \mathbf{w}$, the value for the experimental statistic of the individual significance test with null hypothesis $\beta_{i} = 0$ (in the face of the alternative hypothesis $\beta_{i} \not= 0$, for $i=1,\dots,k$) is:
\begin{equation}
    \label{texp.orto.2}
    t_{exp,O}(\beta_{i}) = \left| \frac{\widehat{\beta}_{i,o}}{\widehat{\sigma}} \right| = \left| \frac{p_{i} \cdot \widehat{\beta}}{\widehat{\sigma}} \right|,
\end{equation}
where $p_{i}$ is the $i$ row of matrix $\mathbf{P}$.

By comparing this expression with the one given in (\ref{texp.orto.1}), it is observed that, as expected, not only the denominator but also the numerator has changed. For this reason, if the null hypothesis is rejected in the initial model, it is not assured that the same occurs in the orthonormal model.

\subsection{A new statistical test to detect multicollinearity}

From the above, the individual significance test from the expression (\ref{texp.orto.2}) is redefined. Thus, the null hypothesis will be rejected, with a significance level $\alpha$, if the following condition is verified:
$$t_{exp,O}(\beta_{i}) > t_{n-k}(1-\alpha/2), \quad i=2,\dots,k,$$
Taking into account the expressions (\ref{texp.orig}), (\ref{redef.VIF}) and (\ref{texp.orto.2}), this is equivalent to
\begin{equation}
    \label{cota.VIF.orto}
    VIF(i) > \left( \frac{t_{n-k}(1-\alpha/2)}{\widehat{\beta}_{i,o}} \right)^{2} \cdot \widehat{var} \left( \widehat{\beta}_{i} \right) \cdot n \cdot var(\mathbf{X}_{i}) = c_{2}(i), \quad i=2,\dots,k.
\end{equation}

Thus, if the $VIF(i)$ is greater than $c_{2}(i)$, the null hypothesis is rejected in the respective individual significance tests in the orthonormal model (with $i=2,\dots,k$). Then, if the null hypothesis is not rejected in the original model and it is verified that $VIF(i) > c_{2}(i)$, it will be possible to conclude that the near-multicollinearity existing in the model is affecting its statistical analysis. In summary, a lower bound for the VIF is established to indicate when the near-multicollinearity is troubling in a manner that can be reinterpreted and presented as a statistical test.

\begin{example}\nonumber
Continuing with data set presented by \cite{Wissell},  Table \ref{tabla.W.orto.2} shows the results of the estimation by OLS of the orthonormal model obtained from the original model. The following conclusions are obtained comparing these results with those shown in Table \ref{tabla.W.orto.1}:
\begin{itemize}
    \item In all cases, the standard deviation has decreased, except for the outstanding consumer credit variable, whose standard deviation has increased.
    \item The values for the experimental statistics of the individual significance tests associated with the intercept and the personal consumption variable have increased, that for the personal income variable has decreased, and that for the outstanding consumer credit variable remains the same. These facts show that the change from the original to the orthonormal model does not guarantee an increased value for the experimental statistic.
    \item The estimation of the coefficient of the personal consumption variable is not significatively different from zero in the original model, but it is in the orthogonal model. Thus, the conclusion will be that the multicollinearity is affecting the statistical analysis of the model. Note that there also is a change in the sign of the estimate, although the goal of the orthogonal model is not to obtain estimates for the coefficients but instead to offer a point of reference against which to measure how much the variances are inflated.
    \item The values corresponding to the estimated variance for the random disturbance, the coefficient of determination and the experimental statistic for the global significance test remain the same.
\end{itemize}

\begin{table}
    \centering
    \begin{tabular}{cccc}
        \hline
        Variable & Estimation & Standard deviation & Experimental $t$ \\
        \hline
        Constant & -27.8823 & 0.9325 & 29.902 \\
        Personal consumption & 11.5925 & 0.9325 & 12.432 \\
        Personal income & -1.3549 & 0.9325 & 1.453 \\
        Outstanding consumer credit & 0.04657 & 0.9325 & 0.499 \\
        \hline
        $\widehat{\sigma}^{2}$ & \multicolumn{3}{c}{$0.9325^{2}$} \\
        $R^{2}$ & \multicolumn{3}{c}{0.9235} \\
        $F_{3,13}$\footnote{Experimental statistic for the global significance test} & \multicolumn{3}{c}{52.3} \\
        \hline
    \end{tabular}
    \caption{OLS estimation for the orthonormal model with data previously presented by  \cite{Wissell}} \label{tabla.W.orto.2}
\end{table}

On the other hand, taking into account the VIF of the independent variables except for the intercept (589.7540, 281.8862 and 189.4874) and their associated bounds (17.80933, 623.1276 and 3545.1672) obtained from expression (\ref{cota.VIF.orto}), only the personal consumption variable verifies that the VIF is higher than the corresponding bound. These results are different from those obtained in Example \ref{ejemplo1}, where the traditional orthogonal model was taken as reference.
\end{example}

\section{A new definition for the VIF and its thresholds: A test for detecting statistical troubling
multicollinearity} \label{new.VIF.orto}

As previously noted, the VIF has traditionally been defined as an increase in the estimated variance of the estimators caused by the degree of multicollinearity existing in the model, taking as reference the variance of the estimators in the orthogonal version of the same model. Using the alternative orthonormal model proposed for  (\ref{model0}) in Section \ref{modelo.orto} and the expression (\ref{redef.VIF}), it is possible to obtain a redefined variance inflation factor, herein named the variance inflation factor (TVIF), as follows:
\begin{equation}
    \label{redef.VIF.bis}
    TVIF(i) = \frac{VIF(i)}{n \cdot var(\mathbf{X}_{i})} = \frac{1}{SSR_{i}}, \quad i=2,\dots,k,
\end{equation}
where $SSR_{i}$ is the sum of the squared residuals of the auxiliary regression (\ref{model_aux}).

On the other hand, given the model (\ref{model0}), the expression deduced for the variance of the estimators (see, for example, \cite{Novales}) is as follows:
$$var (\widehat{\beta}_{i}) = \frac{\sigma^{2}}{SSR_{i}}, \quad i=1,\dots,k.$$
Thus, if it is considered that $var (\widehat{\beta}_{i,o}) = \sigma^{2}$, the following is obtained:
$$\frac{var \left( \widehat{\beta}_{i} \right)}{var \left( \widehat{\beta}_{i,o} \right)} = \frac{1}{SSR_{i}} = TVIF(i), \quad i=1,\dots,k,$$
i.e., in contrast to the traditional VIF, TVIF can be also calculated for $i=1$.

Taking into account the calculation of $SSR_{i}$, TVIF can be expressed as
\begin{equation}
    \label{tvif}
    TVIF(i) = \left( \mathbf{X}_{i}^{t} \mathbf{X}_{i} - \mathbf{X}_{i}^{t} \mathbf{X}_{-i} \cdot \left( \mathbf{X}_{-i}^{t} \mathbf{X}_{-i} \right)^{-1} \cdot \mathbf{X}_{-i}^{t} \mathbf{X}_{i} \right)^{-1}, \quad i=1,\dots,k.
\end{equation}

The TVIF is always positive and also verifies the following:
\begin{itemize}
    \item If the degree of near-multicollinearity is high, then $SSR_{i}$ will be close to zero, and in that case, TVIF will tend to infinity.
    \item If the degree of near-multicollinearity is low, then  $SSR_{i}$ will be high (i.e., its maximum value is $SST_{i}$, as the model always has an intercept), and TVIF will tend to its minimum value  $SST_{i}^{-1}$, where $SST_{i}$ is the sum of the squared totals of the auxiliary regression (\ref{model_aux}).
\end{itemize}

Thus, high values of TVIF are associated with a high degree of multicollinearity. But the question is how high TVIFs have to be to be reflective of troubling multicollinearity. Taking into account the expression (\ref{cota.VIF.orto}), it is possible to conclude that multicollinearity is affecting the statistical analysis of the model if it can be verified that
\begin{equation}
    \label{cota.VIFR}
    TVIF(i) > \left( \frac{t_{n-k}(1-\alpha/2)}{\widehat{\beta}_{i,o}} \right)^{2} \cdot \widehat{var} \left( \widehat{\beta}_{i} \right) = c_{3}(i),
\end{equation}
for any $i=1,\dots,k$.

By following  \cite{OBrien} and taking into account that
$$\widehat{var} \left( \widehat{\beta}_{i} \right) = \widehat{\sigma}^{2} \cdot TVIF(i) = \frac{SSR}{n-k} \cdot TVIF(i),$$
where $SSR$ is the sum of the squared residuals of model (\ref{model0}), there are other factors that counterbalance a high value of TVIF, thereby avoiding high estimated variances for the estimated coefficients. These factors are the SSR and $n$. Thus, an adequate specification of the economic model (i.e., one that implies a good fit and, consequently, a small SSR) and a large sample size can compensate of high TVIF values.
However, in contrast to what occurs in the traditional case, these factor are being considered in threshold $c_{3}(i)$, as established in expression (\ref{cota.VIFR}) in $\widehat{var} \left( \widehat{\beta}_{i} \right)$.

\begin{example}\nonumber
This novel contribution can be illustrated with the data set previously presented by  \cite{KleinGoldberger}, which includes variables for consumption, $\mathbf{C}$, wage incomes, $\mathbf{I}$, non-farm incomes, $\mathbf{InA}$, and farm incomes, $\mathbf{IA}$, in United States from 1936 to 1952, as shown in Table \ref{datos.KG} (data from 1942 to 1944 are not available due to having been war years).

    \begin{table}
        \begin{center}
        \begin{tabular}{ccccc}
            \hline
            Year & $\mathbf{C}$ & $\mathbf{I}$ & $\mathbf{InA}$ & $\mathbf{IA}$ \\
            \hline
            1936 & 62.8 & 43.41 & 17.1 & 3.96 \\
            1937 & 65 & 46.44 & 18.65 & 5.48 \\
            1938 & 63.9 & 44.35 & 17.09 & 4.37 \\
            1939 & 67.5 & 47.82 & 19.28 & 4.51 \\
            1940 & 71.3 & 51.02 & 23.24 & 4.88 \\
            1941 & 76.6 & 58.71 & 28.11 & 6.37 \\
            1945 & 86.3 & 87.69 & 30.29 & 8.96 \\
            1946 & 95.7 & 76.73 & 28.26 & 9.76 \\
            1947 & 98.3 & 75.91 & 27.91 & 9.31 \\
            1948 & 100.3 & 77.62 & 32.3 & 9.85 \\
            1949 & 103.2 & 78.01 & 31.39 & 7.21 \\
            1950 & 108.9 & 83.57 & 35.61 & 7.39 \\
            1951 & 108.5 & 90.59 & 37.58 & 7.98 \\
            1952 & 111.4 & 95.47 & 35.17 & 7.42 \\
            \hline
        \end{tabular}
        \caption{Data set presented by  \cite{KleinGoldberger}} \label{datos.KG}
        \end{center}
    \end{table}
Table \ref{tabla.KG.orto.1} shows the OLS estimations of the model explaining consumption as a function of the rest of the variables. Note that some incoherence is found in relation to the individual significance values of the variables and the global significance of the model.

\begin{table}
    \centering
    \begin{tabular}{cccc}
        \hline
        Variable & Estimator & Standard deviation & Experimental $t$ \\
        \hline
        Constant & 18.7021 & 6.8454 & 2.732 \\
        Wage income & 0.3803 & 0.3121 & 1.218 \\
        Non-farm income & 1.4186 & 0.7204 & 1.969 \\
        Farm income & 0.5331 & 1.3998 & 0.381 \\
        \hline
        $\widehat{\sigma}^{2}$ & \multicolumn{3}{c}{$6.06^{2}$} \\
        $R^{2}$ & \multicolumn{3}{c}{0.9187} \\
        $F_{3,10}$ & \multicolumn{3}{c}{37.68} \\
        \hline
    \end{tabular}
    \caption{OLS estimation of the model previously presented by  \cite{KleinGoldberger}} \label{tabla.KG.orto.1}
\end{table}


The TVIFs are calculated, yielding 1.275947615, 0.002652862, 0.014130621 and 0.053354814, respectively. The associated bounds are also calculated, yielding 0.0021892653, 0.0001206694, 0.0187393601 and 1.8265885762, respectively.

Since the coefficient of the wage income variable is not significantly different from zero, and because it is verified that $0.002652862 > 0.0001206694$, it is concluded that the degree of multicollinearity existing in the model is affecting its statistical analysis.
\end{example}

    Finally, taking into account that in the original model (\ref{model0}), the null hypothesis of the individual significance test is not rejected if:
    $$TVIF(i) > \left( \frac{\widehat{\beta}_{i}}{\widehat{\sigma} \cdot t_{n-k}(1-\alpha/2)} \right)^{2} = c_{0}(i),$$
    whereas in the orthonormal model, the null hypothesis is rejected if $TVIF(i) > c_{3}(i)$. As such, the following theorem can be established:

    \begin{theorem}\label{theorem1}
        Given a multiple linear regression model (\ref{model0}), the degree of near-multicollinearity affects its statistical analysis (with a level of significance of $\alpha\%$) if there is a variable $i$, with $i=1,\dots,k$, that verifies $TVIF(i) > \max \{ c_{0}(i), c_{3}(i) \}$.
    \end{theorem}

    \begin{example}\nonumber
       Tables \ref{teorema.w} and \ref{teorema.kg} present the results of applying Theorem \ref{theorem1} to the \cite{Wissell} and \cite{KleinGoldberger} models, respectively.  Note that in both cases, there is a variable $i$ verifying that $TVIF(i) > \max \{ c_{0}(i), c_{3}(i) \}$, and consequently, we can conclude that the degree of near-multicollinearity is affecting the statistical analysis in both models (with a level of significance of $\alpha\%$). \hfill $\Box$
        \begin{table}
            \centering
            \begin{tabular}{ccccc}
                \hline
                $i$ & $TVIF(i)$ & $c_{0}(i)$ & $c_{3}(i)$ & Affects \\
                \hline
                1 & 194.8661 & 7.371069 & 1.017198 & Yes \\
                2 & 30.32628 & 4.456018 & 0.9157898 & Yes \\
                3 & 4.765888 & 2.399341 & 10.53598 & No \\
                4 & 0.00003821626 & 0.000002042640 & 0.0007149977 & No \\
                \hline
            \end{tabular}
            \caption{Theorem \ref{theorem1} results of the model previously presented by \cite{Wissell}} \label{teorema.w}
        \end{table}
        \begin{table}
            \centering
            \begin{tabular}{ccccc}
                \hline
                $i$ & $TVIF(i)$ & $c_{0}(i)$ & $c_{3}(i)$ & Affects \\
                \hline
                1 & 1.275947615 & 1.9183829079 & 0.0021892653 & No \\
                2 & 0.002652862 & 0.0007931658 & 0.0001206694 & Yes \\
                3 & 0.014130621 & 0.0110372472 & 0.0187393601 & No \\
                4 & 0.053354814 & 0.0015584988 & 1.8265885762 & No \\
                \hline
            \end{tabular}
            \caption{Theorem \ref{theorem1} results of the model previously presented by \cite{KleinGoldberger}} \label{teorema.kg}
        \end{table}
    \end{example}

\section{Threshold for Stewart's index} \label{cotas.ki.apartiTVIFR}

 \cite{Stewart:1987} presented the index $S_{i}^{2}$, which measures the relation between column $i$ of matrix $\mathbf{X}$ and the rest of the columns of $\mathbf{X}$ of model (\ref{model0}) from the following expression:
$$S_{i}^{2} = \frac{\mathbf{X}_{i}^{t} \mathbf{X}_{i}}{ \mathbf{X}_{i}^{t} \mathbf{X}_{i} - \mathbf{X}_{i}^{t} \mathbf{X}_{-i} \cdot \left( \mathbf{X}_{-i}^{t} \mathbf{X}_{-i} \right)^{-1} \cdot \mathbf{X}_{-i}^{t} \mathbf{X}_{i}}.$$

For $i \geq 2$, it can be found (see  \cite{Salmeron2019b}) that:
$$S_{i}^{2} = VIF(i) + n \cdot \frac{\overline{\mathbf{X}}_{i}^{2}}{SCR_{i}},$$
where $\overline{\mathbf{X}}_{i}$ is the mean of the variable $i$ of $\mathbf{X}$, and $SSR_{i}$ is the sum of the squared residuals of the auxiliary regression (\ref{model_aux}).

Although Stewart identified this index with the VIF, both measures only coincide if the variable where they are calculated has a mean of zero (see \cite{Salmeron2019b} for more detail). Another interesting aspect that makes the $S_{i}^{2}$ measure different from the VIF is that it can be obtained for $i=1$ (i.e., for the intercept). In this case, it is possible to diagnose the non-essential multicollinearity (i.e., the relationship between the intercept and the rest of the independent variables of the model) existing in the model. In addition, \cite{SalmeronARXIV} showed that this index coincides with the VIF in models where there is no intercept. This fact justifies the relevance of the following development.

Thus, from (\ref{tvif}), it is possible to obtain the following expression for Stewart's index:
\begin{equation}
    S_{i}^{2} = TVIF(i) \cdot \mathbf{X}_{i}^{t} \mathbf{X}_{i} = TVIF(i) \cdot \sum \limits_{j=1}^{n} X_{ji}^{2}, \quad i=1,\dots,k. \label{ki.VIFR}
\end{equation}
Note that when the observations are expressed in unit length, it is verified that $|| \mathbf{X}_{i} ||^{2} = \sum \limits_{j=1}^{n} X_{ji}^{2} = 1$ and, then, $S_{i}^{2} = TVIF(i)$.

If in the expression (\ref{cota.VIFR}), it is established that the values of TVIF are troubling when the condition
$$TVIF(i) > \left( \frac{t_{n-k}(1-\alpha/2)}{\widehat{\beta}_{i,o}} \right)^{2} \cdot \widehat{var} \left( \widehat{\beta}_{i} \right), \quad i=1,\dots,k$$
is verified, then from the relation given in (\ref{ki.VIFR}), it will similarly be verifiable that the values of $S_{i}^{2}$, with $i=1,\dots,k$, affect the statistical analysis of the model, using the following condition for verification:
\begin{equation}
    \label{cotas.ki}
    S_{i}^{2} > \left( \frac{t_{n-k}(1-\alpha/2)}{\widehat{\beta}_{i,o}} \right)^{2} \cdot \widehat{var} \left( \widehat{\beta}_{i} \right)  \cdot \sum \limits_{j=1}^{n} X_{ji}^{2}, \quad i=1,\dots,k.
\end{equation}

Thus, we have established thresholds for Stewart's index (for $i=1,\dots, k$) from which we can conclude that the multicollinearity  is affecting to the statistical analysis of the model.

\section{Conclusions} \label{conclusiones.VIF}

Traditionally, the Variance Inflation Factor has been defined as increases in the estimated variances of the estimated coefficients of an econometric model caused by the degree of multicollinearity existing in the model, taking as reference the orthogonal version of the same model.

In this paper, it is shown that the orthogonal model traditionally considered is not an adequate point of reference. It also proposes an alternative orthogonal model than leads to a lower bound for the VIF that will indicate if the degree of multicollinearity existing in the model affects its statistical analysis. These thresholds serve as complements to the results presented by \cite{OBrien}, who stated that the estimated variances depend on other factors that can counterbalance a high value of the VIF---for example, the size of the sample or the estimated variance of the independent variables. Thus, the thresholds presented for the VIF also depend on these factors’ meeting a threshold associated with each independent variable (except for the intercept). Note that these thresholds will indicate whether the degree of multicollinearity affects the statistical analysis.

On the other hand, parting from the traditional definition of the VIF, this paper presents the redefined variance inflation factor (TVIF). TVIF coincides with the inverse of the sum of the squared residuals of the auxiliary regression of the independent variable $i$ as a function of the rest of the independent variables of the model. Thresholds for this measure are also provided to determine when the values of TVIF indicate that the degree of multicollinearity in the model affects its statistical analysis. Thresholds for the Stewart's index are also presented, since it is related to TVIF.

Finally, this paper also presents a statistical test to determine if the degree of multicollinearity existing in the model affects its statistical analysis. This analytic tool allows investigators to conclude whether the degree of multicollinearity is statistically troubling and whether it is necessary to treat it. We consider this to be a relevant contribution since, to the best of our knowledge, the sole extant example of such a measure, that presented by  \cite{FarrarGlauber}, has been strongly criticized (see, for example, \cite{CriticaFarrar1}, \cite{CriticaFarrar2}, \cite{CriticaFarrar3} and  \cite{CriticaFarrar4}); consequently, this new test will fill a gap in the scientific literature.

\bibliographystyle{chicago}
\bibliography{bib}

\end{document}